# ACSYS IN A BOX

C. Briegel, D. Finstrom, B. Hendricks, C. King, S. Lackey, R. Neswold, D. Nicklaus, J. Patrick, A. Petrov, R. Rechenmacher, C. Schumann, J. Smedinghoff

FNAL[†], Batavia, IL 60510, U.S.A.

*Abstract*
   The Accelerator Control System at Fermilab has evolved to enable this relatively large control system to be encapsulated into a "box" such as a laptop. The goal was to provide a platform isolated from the "online" control system. This platform can be used internally for making major upgrades and modifications without impacting operations. It also provides a standalone environment for research and development including a turnkey control system for collaborators. Over time, the code base running on Scientific Linux has enabled all the salient features of the Fermilab's control system to be captured in an off-the-shelf laptop. The anticipated additional benefits of packaging the system include improved maintenance, reliability, documentation, and future enhancements.

## INTRODUCTION

The Fermilab control system [1] [2] has never been packaged for utilization by other institutions. When it was needed at remote sites, a special effort was made to provide an autonomous system with an umbilical cord to Fermilab to ease maintenance and operation. Until recently, there was never a need for a stand-alone implementation at an isolated site. New facilities [3] for accelerator R&D, collaboration with other institutions, remote experimental sites, and the desire to test new functionality in an isolated environment suggested the demand for packaging the Fermilab control system. The package with evolving enhancements is known as ACSys (Accelerator Control System) and has been demonstrated to run on a laptop.

## REQUIREMENTS

This implementation is not intended to run on all operating systems and processors. The approach is to build for a rich environment while keeping the configurations manageable. Thus, a laptop running Scientific Linux is the foundation for a "single-box" solution. Additional "boxes" can be added as needed to interface hardware, increase the capability or isolate systems utilizing Scientific Linux or a vxWorks framework. Currently, Sybase is the supported database architecture.

## PROCESS

The original goal was to provide a minimal control system implementing only the bare essentials. This reduced system was built on a Fermilab "standard-issue" laptop utilizing a scripting language, a device-only database, and a local front-end for data acquisition. While this version brought in much of the infrastructure, it left many desirable features out of the implementation.

The gradual extension of the goals evolved into a full-featured control system analogous to the existing Fermilab control system. The work to minimize the system proved to be more time-consuming in the long run and was not capable of maximizing the potential of the system. A full implementation enabled a platform for testing enhancements and provided an autonomous control system at remote locations.

ACSys is effectively the Fermilab control system with the ability to redirect the resources to coincide in a single instance. For protection of the existing control system, a unique IP port was used for communication. Also, distinct multicast addresses were used for various services such as state transitions and event notification.

## COMPONENTS

ACSys is a tiered implementation (refer to Fig. 1.) It consists of a console tier for applications and displays; a central services tier for database, logging, save/restore, data consolidation and alarms; and a front-end tier for data access and control. These tiers are logical and can be associated with one or more physical pieces of hardware.

ACSys is a layered protocol (refer to Fig. 1.) Their is a peer-to-peer messaging protocol tunnelled on UDP/IP. This protocol is then used by several data acquisition protocols for reading and setting data, plotting data in real-time, and triggered data snapshots. Alarm handling involves another protocol providing unsolicited event and exception notifications.

Implementing layered protocols enables many implementations to coexist. For instance, data acquisition can be accomplished with multiple protocols in the same architecture. This enables the adiabatic evolution of enhancements without imposing change to stable or critical systems.

### Console Tier

The console is a X-based server/client of displays and a client for data manipulation. These displays can be implemented in C++, Java, a drag-and-drop graphical display manager (Synoptic [4]), Web-based, or with a scripting language (ACL). The console optimizes data

---

acquisition with a data pool manager for several console instances.

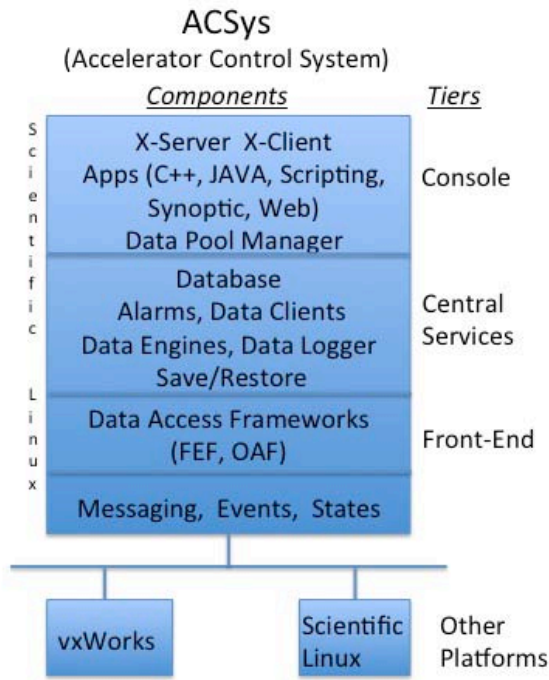

Figure 1: ACSys Configuration.

*Central Services Tier*

The database is a key component of the central tier. Sybase has been utilized for many years and the system capitalizes on several Sybase features. These features have been enumerated and ACSys provides an excellent development platform for an alternative database in the future.

The alarms server is a central repository for all announced alarms. These alarms are then distributed to consoles for individually configured displays. The monitoring of alarm limits is performed by a front-end. An unsolicited message is sent to the server when the condition becomes either bad or good.

The central tier provides a data engine to efficiently acquire and distribute data to clients such as save/restore. These engines also provide the equivalent data pool manager for Java applications.

Data loggers also use the engines to acquire a list of devices based on time or event. The data logger provides an archival circular repository. The duration of the repository is based on frequency and size of the data.

*Front-end Tier*

Prior to this effort, a new front-end framework (FEF) was being developed. This framework is designed for Scientific Linux. The framework is written in Erlang with C++ interfaces and is functional. A more mature framework called OAF (Open Access Frontend) is written in Java and has been used primarily for software devices or calculations.

The FEF [5] and OAF architectures complement other frameworks operating in vxWorks. These environments can be added to the configuration with additional platforms (outside the box).

*Global Services*

Other pieces of the infrastructure exist in ACSys. A peer-to-peer messaging system provides protocol payloads between cooperating connected tasks. Also, by setting a registered device, a global state transition is announced over a multicast frame. Normally, a hardware clock synchronizes data and its collection by asserting events in a serial protocol. A software solution can provide similar functionality generating periodic, sequenced and one-shot events. Global event notification occurs via a multicast frame for either hardware or software generated events.

The components listed above are not an exhaustive list of all available implementations, but are distinguished as core components. These components are readily available at this time to the implementer.

## RESULTS

This effort has resulted in a fully capable stand-alone control system. ACSys provides a collaboration infrastructure consistent with the global control system at Fermilab. ACSys provides a testing environment for control system and accelerator system development. ACSys can also provide an isolated control system for remote operation.

A potential candidate for adopting the control system is an evolving experiment at Fermilab. In this case, the experiment has an on-site and remote detector. ACSys has positioned itself as a viable system to support this experiment. The control system offers common support and development with Fermilab personnel at a time of dwindling resources.

Current R&D efforts at Fermilab are using the Fermilab control system. A successful collaboration should ensure future instrumentation and analysis can be accomplished with Fermilab supported tools. ACSys makes remote R&D for collaborators a reality. This reality will aid in integration into the on-site control system providing a more maintainable system.

There are other benefits. Developing a generic system forces an understanding of dependencies that are sometimes taken for granted. These dependencies evolve since our target is not to commercialize the system, but to build a system for our users at Fermilab. Targeting other uses forces a reflection on implementation choices.

There are future benefits. The ACSys meetings and discussions are useful to discuss new directions and development. The direction of the group is providing input to new requirements of the control system. Hopefully, this direction will evolve into an active global development for the control system. The packaging of a system implies the documentation must be enhanced. This additional requirement is always a desirable bonus.